\documentclass[12pt]{iopart}

\usepackage{natbib}
\usepackage{graphicx}
\usepackage{comment}
\usepackage[colorinlistoftodos]{todonotes}
\usepackage{caption}
\usepackage{subcaption}
\usepackage{siunitx}
\usepackage{lineno}
\usepackage[normalem]{ulem}
\usepackage[utf8]{inputenc}
\usepackage[T1]{fontenc}

\begin{document}

\title[Deep learning in thermoluminescence dosimetry]{No more glowing in the dark: How deep learning improves exposure date estimation in thermoluminescence dosimetry
}%\todo{ON: I think we should change the title}

% authors as required in guidelines 3.2
\author{F.~Mentzel$^1$, E.~Derugin$^1$, H.~Jansen$^1$, K.~Kröninger$^1$, O.~Nackenhorst$^1$,
%	M.~Rempe$^1$,
J.~Walbersloh$^2$ and J.~Weingarten$^1$
}

\address{$^1$TU Dortmund University, August-Schmidt-Straße 1, 44227 Dortmund, NRW, Germany}
\address{$^2$Materialprüfungsamt NRW, Marsbruchstraße 186, 44287 Dortmund, NRW, Germany}
\ead{\mailto{florian.mentzel@tu-dortmund.de}}
\vspace{10pt}
\begin{indented}
	\item[]May 2021
\end{indented}

\begin{abstract} % currently 150 words
The time- or temperature-resolved detector signal from a thermoluminescence dosimeter can reveal additional information about circumstances of an exposure to ionizing irradiation.
We present studies using deep neural networks to estimate the date of a single irradiation with 12 mSv within a monitoring interval of 42 days from glow curves of novel TL-DOS personal dosimeters developed by the Materialprüfungsamt NRW in cooperation with TU Dortmund University.
Using a deep convolutional network, the irradiation date can be predicted from raw time-resolved glow curve data with an uncertainty of roughly 1-2 days on a 68\% confidence level without the need for a prior transformation into temperature space and a subsequent glow curve deconvolution. This corresponds to a significant improvement in prediction accuracy compared to a prior publication, which yielded a prediction uncertainty of 2-4 days using features obtained from a glow curve deconvolution as input to a neural network.
\end{abstract}%\todo{I think we should work a bit on the abstract in order to be more consistent with our story line}

\vspace{2pc}
\\
\submitto{\JRP}
\maketitle

\clearpage

%\begin{linenumbers}
\section{Introduction}
%Ionizing irradiation is necessary for many technological advances, e. g. in electricity generation, material testing, food safety as well as in medical imaging or cancer treatment \citep{personal dose monitoring in material testing, ppm in nuclear energy, ppm in food safety, ppm in medical imaging}.
%Although these applications benefit from ionizing radiation its potential harm can be severe, when humans are exposed to it, ranging from acute radiation illness in cases of strong exposure to elevated risk of cancer even at minimal exposure \citep{cancer risk, cancer risk at low exposures}. For this reason, the use of ionizing radiation and its sources is strictly regulated, and radiation protection concepts are enforced for the safety of people handling it and those being exposed to during medical treatments \citep{Strahlenschutzverordnung, Roetgenverordnung}. An important measure to supervise the effectiveness of radiation protection
%concepts for occupationally exposed people is the personal dose monitoring \citep{importance of dose monitoring for safety}.
%\\
The goal of routine personal dose monitoring is to quantify the exposure to ionizing irradiation of a person within a given monitoring interval, most commonly one month \citep{SSK.AnforderungenPersonendosimeter}. Thereby, the energy deposited by radiation in the persons tissue, leading to potential health hazards, can be estimated. Upon detection of an excess in exposure to radiation, the radiation protection officer needs to determine its source and whether the radiation protection concept needs to be improved \citep{SSK.VoraussetzungenStrahlenschutz}. This time-consuming task can be facilitated using active dosimeters which alert the user immediately in case of an exposure \citep{Ciraj-Bjelac.2018, Conti.2013}. However, due to technical reasons, which are mostly related to pulsed radiation fields \citep{Ambrosi.2009}, official dose monitoring is usually performed using passive dosimeters often based on the effects of optically stimulated luminescence (OSL) \citep{Yukihara_2008,SOMMER20111818,SHOLOM201433} or thermoluminescence (TL) \citep{McKeever.1995,Kron.2021}.
Active dosimeters can therefore in many cases only be carried as an additional, secondary device resulting in higher costs and inconvenience as a person then has to carry two dosimeters at a time.
\\\\
In our studies we use TL-DOS thermoluminescence dosimeters \citep{Walbersloh.RadProt.2016} which are developed in a joint research effort of one of the official dose monitoring agencies in Germany, the Materialprüfungsamt NRW (MPA NRW), and TU Dortmund University.
The capability of the TL-DOS system to reliably estimate the total received dose within the monitoring for gamma \citep{Walbersloh.RadProt.2016,Theinert.RadMeas.2017} and in the future potentially also neutron irradiation \citep{Heiny2020} has been demonstrated in several past studies. More recent studies show that in addition to the total received dose, the analysis of the thermoluminescence signal from the detectors, the so-called glow curves, can provide additional information about the circumstances of the detector irradiation. Most recently, the date of a single exposure within a monitoring interval of $42\,$days was reconstructed with a prediction uncertainty of 2-4 days using neural networks \citep{Theinert.PhD.2018,Kroeninger.RadMeas.2019,Mentzel.RadMeas.2020}.
\\
The possibility to estimate the date of an exposure from the signal of a passive dosimeter can help a radiation protection officer in charge identifying the event that led to the exposure. This can facilitate the workflow on reporting and potentially improving the radiation protection programme in place without the need for additional active dosimeters. While the exposure date estimation is still in a proof-of-concept phase, other aspects like the differentiation between a single and continuous exposure, indicating an accident or a faulty radiation protection setup, have been found to show potential for research and are currently investigated.
\\\\
Those prior works rely on extensive preprocessing of the glow curves and feature-engineering of variables that can be used as input to the neural networks.
This includes a transformation of the glow curves from a time to a temperature space \citep{Theinert.RadMeas.2017}, and a subsequent deconvolution of the glow curve in individual peaks in order to obtain meaningful features.
While glow curve models and fit procedures are formulated in the temperature domain~\citep{Kitis.2006,Sabini.NIM.2002,ELIYAHU2017282,ELIYAHU2014600}, the TL-DOS readout automate however records the glow curves as photon counts per time interval. The current automate design is optimized for high sensitivity and detector throughput using a hot plate for heating, which does not allow for a reliable direct temperature measurement.	
As small uncertainties deviations in the transformation step can lead to large uncertainties on the subsequent deconvolution, this step was found to be a driving factor of uncertainties in the previous studies~\citep{Theinert.PhD.2018,Kroeninger.RadMeas.2019,Mentzel.RadMeas.2020}.
In addition, the underlying physical model of the glow curve deconvolution itself is still a field of open research \citep{ELIYAHU2017282,ELIYAHU2014600}, making the development of robust fit routines for glow curves resulting from various irradiation scenarios(e.g. multiple irradiations or an irradiation with different sources) an even more challenging task.
\\
This paper presents two approaches to glow curve analysis without the need for a transformation into temperature space and a subsequent glow curve deconvolution. The first approach uses glow curve descriptors derived directly from the raw time-dependent signal as input to a deep neural network in order to reduce uncertainties arising form the model-dependent temperature reconstruction.
In the second approach, the raw time-dependent signal itself is passed to a convolutional neural network in order to train convolutional filter, which create glow curve features without human knowledge. The motivation behind this approach is that any knowledge-based feature engineering implies dimensionality reduction, which leads to a loss of information content in the raw data.
Both approaches require neither extensive preprocessing of the data nor the time-consuming deconvolution into individual peaks, which additionally requires the assumption of an underlying model of the TL process. Thus, they are more suitable for a fast and robust TL-DOS readout system.
\\\\
This paper is structured as follows: In \sref{section:Methods}, the TL-DOS system is briefly introduced, followed by a description of the recorded data which is used in the presented studies and a brief introduction of the used machine learning algorithms.
Subsequently, the results of the irradiation date predictions using the the two new deep learning approaches are presented in \sref{section:Results} and compared to the previously reported results in \sref{section:Discussion}.
\Sref{section:Conclusion} summarises the findings and draws conclusions.

\section{Methods \label{section:Methods}}

\subsection{The TL-DOS system \label{section:Methods:TLDOS}}
\begin{figure}[!ht]
	\begin{center}
	 \begin{subfigure}[b]{0.48\textwidth}
		\centering
		\includegraphics[width=\textwidth]{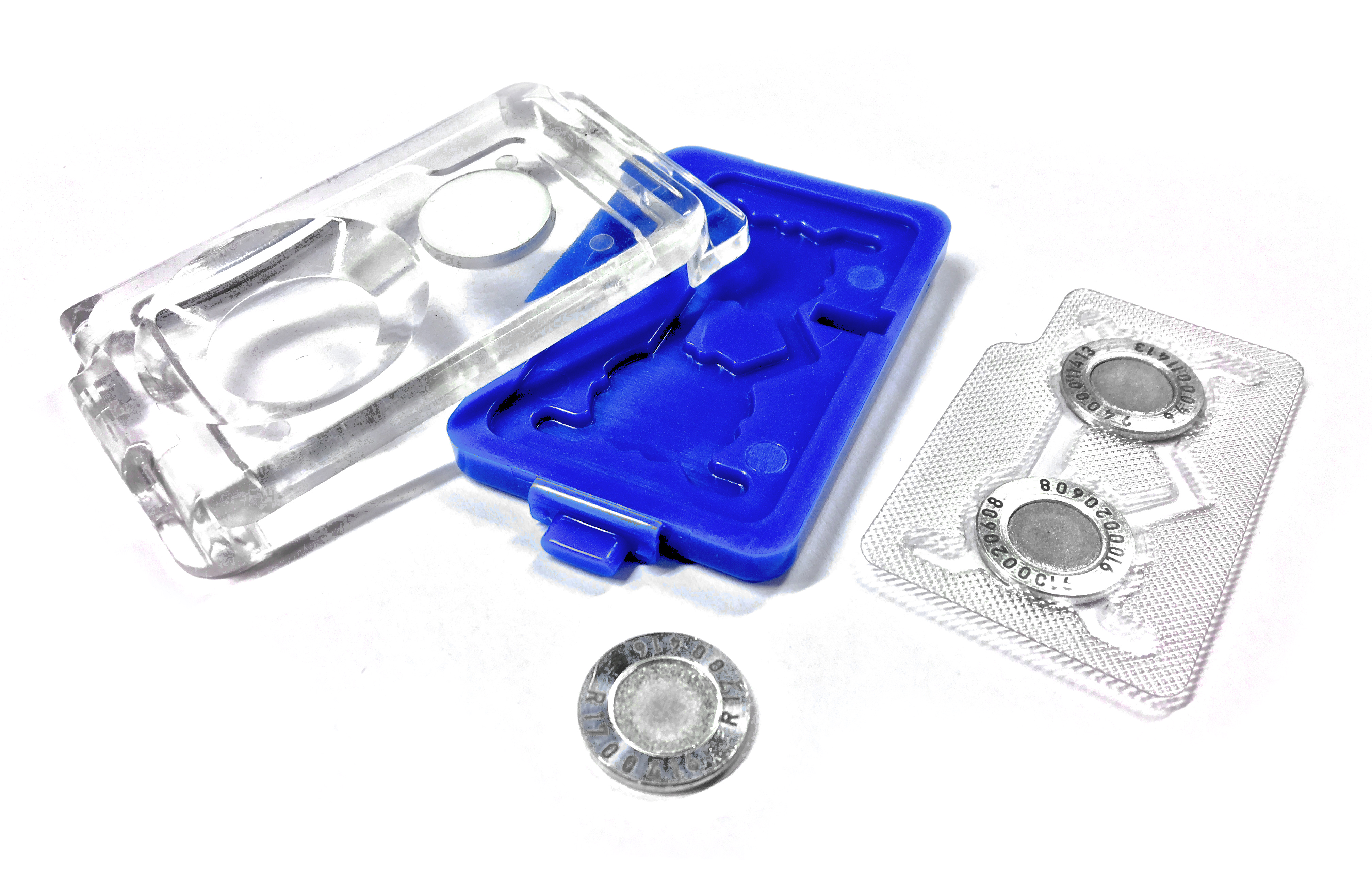}
		\caption{TL-DOS cassette.}
		\label{fig:tl-dos:badge}
	\end{subfigure}
	\hfill
	\begin{subfigure}[b]{0.48\textwidth}
		\centering
		\includegraphics[width=\textwidth]{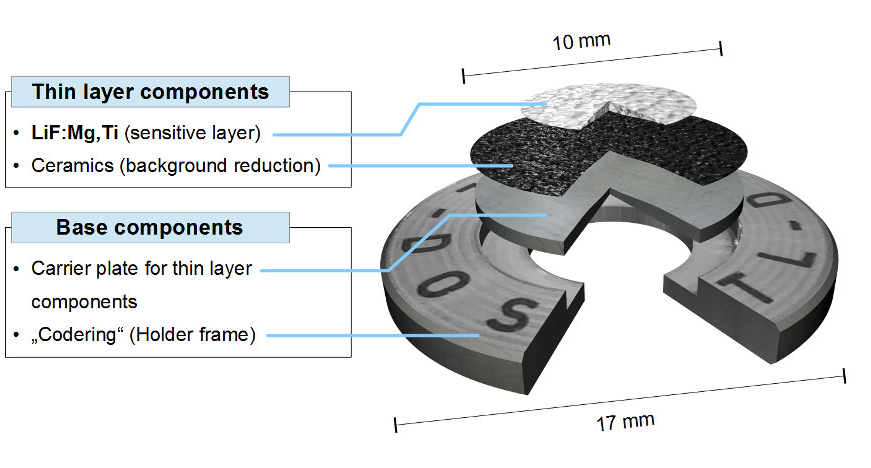}
		\caption{TL-DOS detector.}
		\label{fig:tl-dos:detector}
	\end{subfigure}
	\caption{A picture of the (a) TL-DOS dosimeter cassette (image courtesy: MPA NRW) and its cover with two blistered detectors and a single detector and a schematic of a single (b) TL-DOS detector as used inside the cassette~\citep{Theinert.RadMeas.2017}.}
	\end{center}
\end{figure}

\begin{figure}[h]
	\centering
	\includegraphics[width=0.5\linewidth]{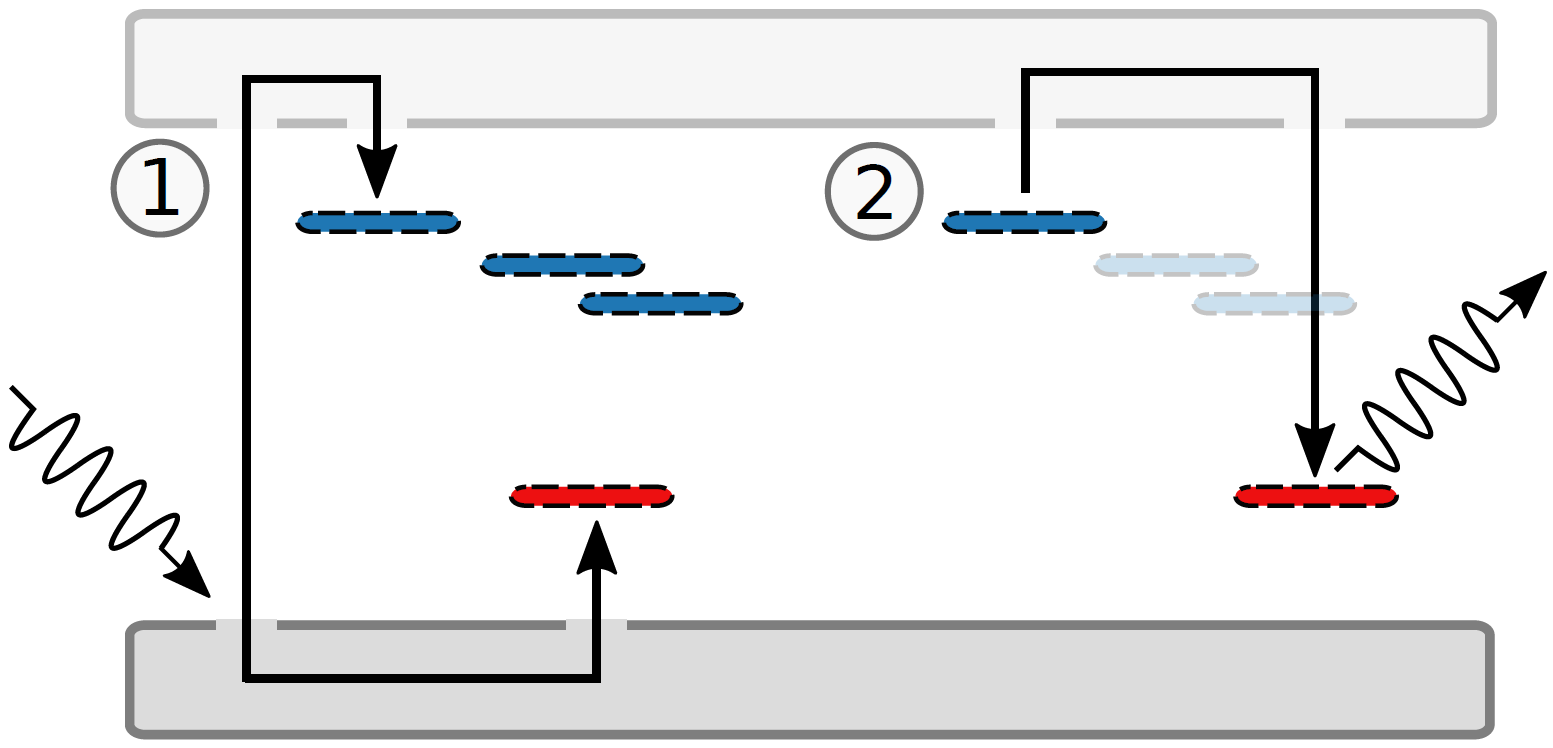}
	\caption{
		Schematic of the thermoluminescence process using the band model for crystals~\citep{Theinert.PhD.2018}.
		(1): Ionizing irradiation creates electron hole pairs in the valence band (dark grey, bottom).
		Excited electrons are elevated into the conduction band (light grey, top) and trapped in metastable states (blue).
		(2): upon heating, trapped electrons are released from the metastable states and recombine with holes (red) under emission of photons.
	}
	\label{fig:tl-dos:tl-process-schematic}
\end{figure}

\begin{figure}[!ht]
	\centering
	\includegraphics[width=0.6\linewidth]{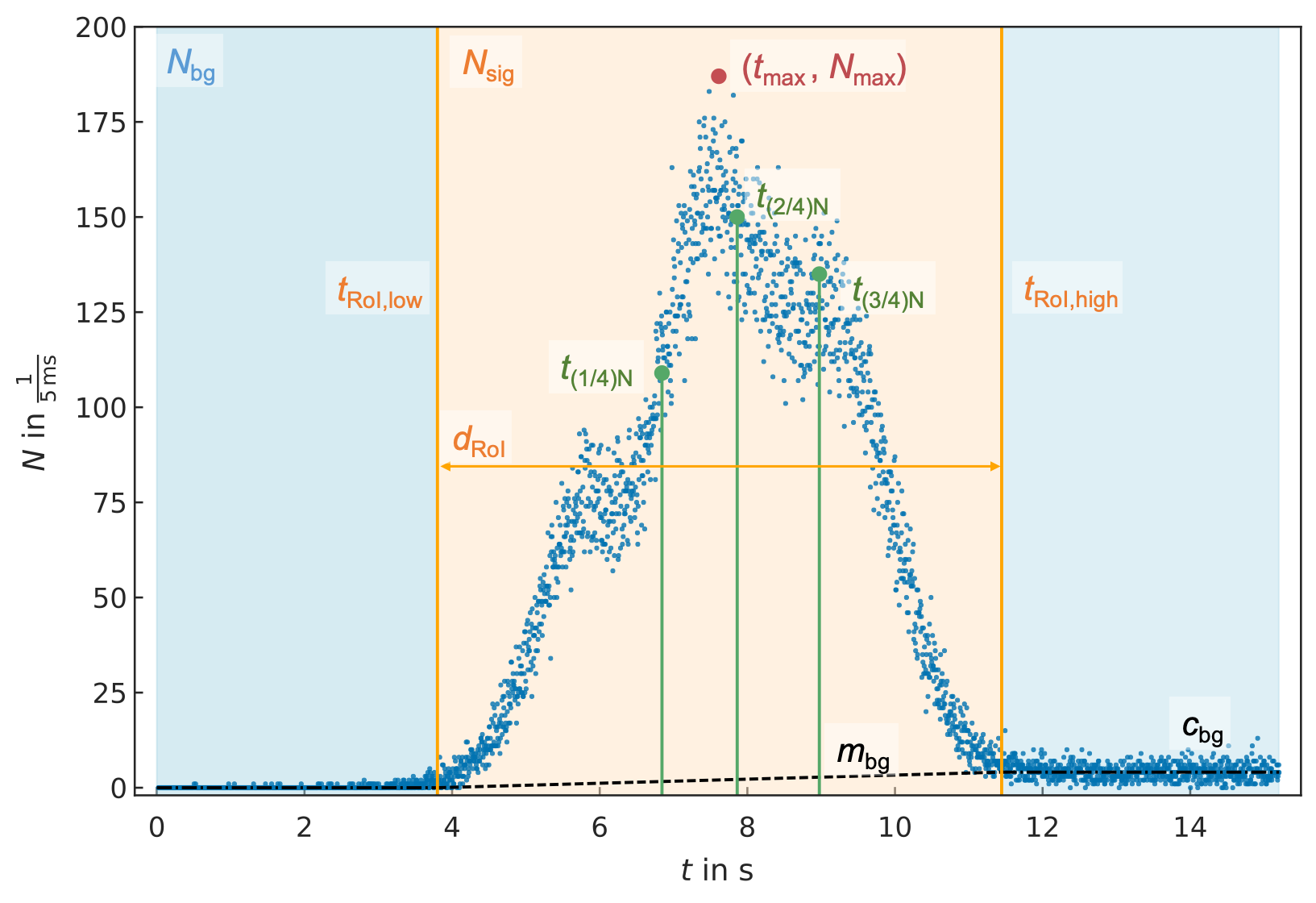}
	\caption{Illustration of a glow curve with the corresponding glow curve parameters in time space.}
	\label{fig:tl-dos:time-glow-curve}
\end{figure}

The thermoluminescence dosimetry (TL-DOS) system is developed by the MPA NRW and TU Dortmund University to replace the film badge dosimeters currently used in personal dose monitoring in parts of Germany.
The dosimeter cassette is shown in \fref{fig:tl-dos:badge}.
Each cassette contains two of the detectors, as depicted as a schematic in \fref{fig:tl-dos:detector}, which can be either used for redundancy or to measure an additional dosimetric quantity such as the skin dose $H_\mathrm{p}(0.07)$ using a different filter in front of the detector chip.
\\
Each detector core is pressed into an aluminium ring with identification codes.
The detector cores consists of a thin layer of Lithium Fluoride (LiF) doped with Magnesium (Mg) and Titanium (Ti) as detecting material. This sensitive layer is sintered onto an aluminium carrier plate, which is covered by a flame-sprayed layer of ceramics serving as background reduction.
The used detector material \hbox{LiF:Mg, Ti} is one of the most commonly used thermoluminescence material for dosimetry \citep{Kortov2007}.
\\
The thermoluminescence mechanism can be explained using the band model for crystals (e.g. \citep{Randall1945,Randall1945-2}), of which a schematic is shown in \fref{fig:tl-dos:tl-process-schematic}:
Thermoluminescent materials are isolators that have metastable, localized states within their band gap.
When exposed to ionizing irradiation with higher energy than the band gap, the irradiation can excite electrons from the valence to the conduction band.
The free remaining spot in the valence band is referred to as hole and acts like a positive charge carrier.
The free electron and the hole can move in the crystal lattice until they either recombine with each other or another respective charge carrier under photon emission, or are captured by metastable states (traps) in the band gap, which arise from impurities like dopant atoms with an excess or a shortage of valence electrons.
Positively charged states can capture electrons, negatively charged states capture holes.
\\
\\
The thermoluminescence emission process is complicated and is explained in detail in the literature, e.g. \citep{horowitz2020thermoluminescence2}. In a simplified model, the kinetic energy of trapped charge carriers rises when heating the material, until it is sufficient for the carriers to be released from the trap state. 
As a result, they can then move around freely again until they recombine under photon emission.
The number of emitted photons in dependence of the material temperature or the measurement time is referred to as a glow curve.
\\
An exemplary single glow curve in the time domain as recorded using the TL-DOS readout system showing the registered photon counts per $\SI{5}{\milli\second}$ count interval is displayed in \fref{fig:tl-dos:time-glow-curve}.
The number of photons emitted within a region of interest (RoI) can be used to estimate the energy deposited by the ionizing irradiation.
Multiple other features can be derived from the shape of the glow curve as depicted in the figure.
The features used in the scope of this study are described in the following. The total and average number of counted photons are denoted as $N_\mathrm{tot}$ and $N_\mathrm{mean}$, respectively. The maximum of the glow curve is reached after $t_\mathrm{max}$ seconds with $N_\mathrm{max}$ photons per count interval.
The times $t_\mathrm{RoI, low}$ and $t_\mathrm{RoI, high}$ mark the start and end of the RoI respectively while $d_\mathrm{RoI}$ denotes its duration.
At the timestamps $t_\mathrm{(i/4)N}$ with i=(1,2,3), the cumulative sum of the photons reaches the fraction i/4 of the total count of signal photons $N_\mathrm{RoI}$ within the RoI. The average number of events occurring after the region of interest is denoted as $c_\mathrm{bg}$. Photon counts outside the RoI or below a dynamic threshold connecting the points $(t_\mathrm{RoI, low}, 0)$ and $(t_\mathrm{RoI, high}, c_\mathrm{bg})$ are considered to be the background $N_\mathrm{bg}$. The slope of this linear function is denoted with $m_\mathrm{bg}$. The difference between the total number of counts and background counts is the signal count $N_\mathrm{sig}$.
\\

\subsection{Dataset for irradiation date estimation \label{section:Methods:dataset}}
\begin{figure}
	\centering
	\includegraphics[width=0.6\linewidth]{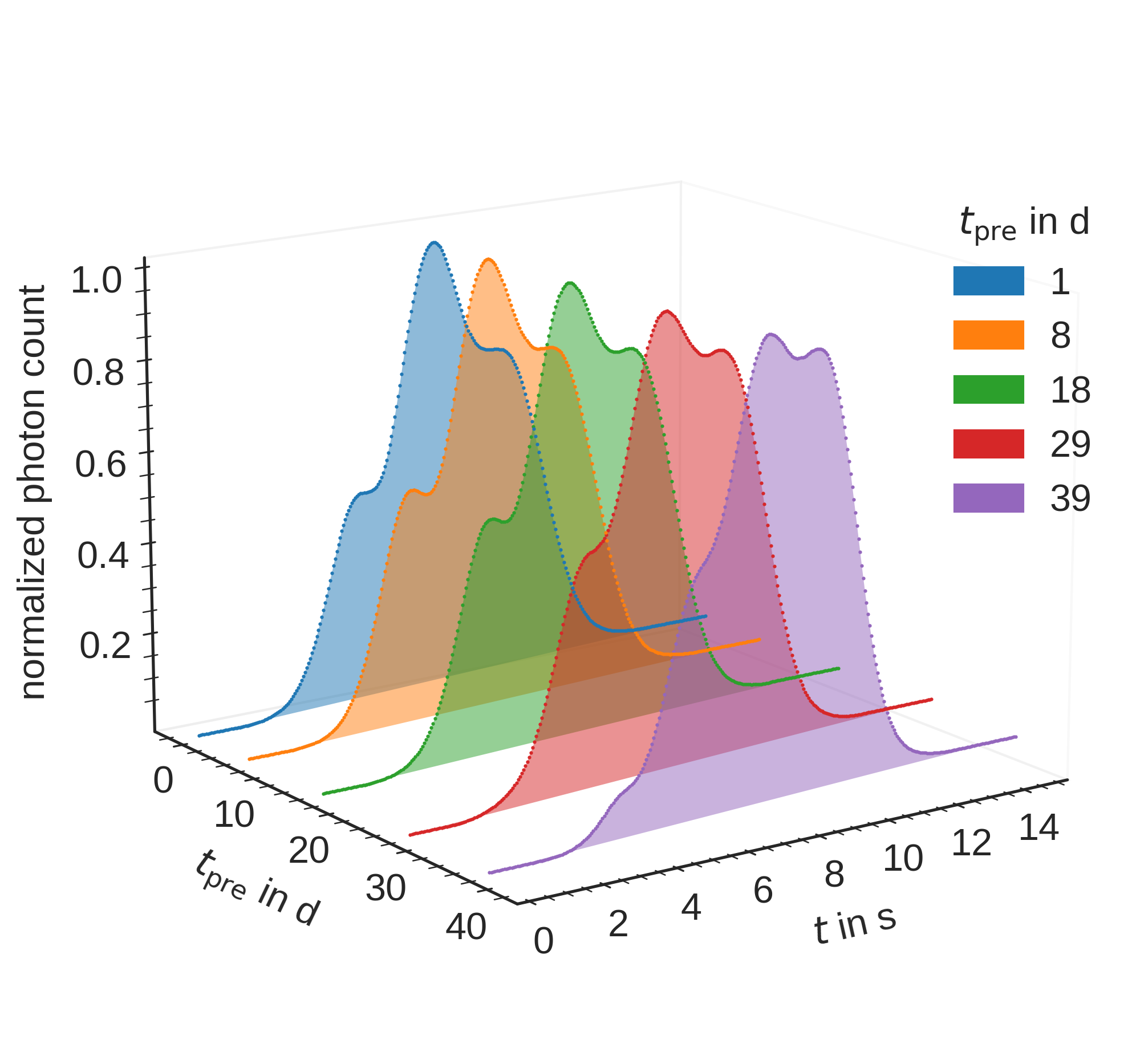}
	\caption{Glow curve shape in dependence of the pre-irradiation storage time $t_\mathrm{pre}$ normalized by the glow curve maximum.}
	\label{fig:gcana:3dfadingcurves}
\end{figure}

The measured data presented in the following were used for the first time in \citep{Mentzel.RadMeas.2020} for a glow curve analysis.
\\
The data set was obtained with 548 TL-DOS detectors, which were prepared for usage on the same day, split into groups of 40-50 detectors (except for one outlier group with 17 detectors due to technical problems) and stored at constant room temperature of $T_0=20^\circ$C.
Afterwards, the detectors were irradiated with a single irradiation dose of $D=\SI{12}{\milli\sievert}$ using a Cs-137 irradiation source ($E = \SI{662}{\kilo\electronvolt}$) on twelve different dates ($t_\mathrm{pre} = [1, 4, 8, 11, 15, 18, 22, 25, 29, 32, 36, 39]\,$d), where the pre-irradiation storage time $t_\mathrm{pre}$ is the number of days between detector preparation and irradiation.
The relatively high dose of $D=\SI{12}{\milli\sievert}$ was chosen to achieve a large number of thermoluminescence photons to facilitate this proof-of-concept.
\\
After irradiation, they were again stored at constant room temperature and read out after the monitoring interval of 42 days counting from the day of the preparation.
\Fref{fig:gcana:3dfadingcurves} shows the averaged glow curves for five different exemplary pre-irradiation storage times $t_\mathrm{pre}$ normalized by their respective maximum.
The blue curve is obtained from detectors which were irradiated one day after the detectors were prepared, while the purple curve is obtained from detectors which were irradiated three days before readout. The orange, green and red curves in between have an intermediate ratio of pre- to post-irradiation storage time.
\\
While the figure suggests an increase of the rightmost peak, this is in fact a result of the decrease of the central peak which is used for the normalization of the curve with increasing pre-irradiation storage time $t_\mathrm{pre}$. This can be seen more intuitively in \citep{Mentzel.RadMeas.2020}.
For the training of the neural networks, only 70\% of the measured glow curves are used, which is referred to as \textit{training data}. 
The remaining 30\% of the data (\textit{test data}) is not used during the training nor the model selection (hyperparameter search), but only to obtain an unbiased performance evaluation of the final model.
The split is performed in such a way, that the fraction of samples of each measurement group is the same in the training and test data set.

\subsection{Machine learning models for irradiation date estimations \label{section:Methods:ML}}

The first applications of machine learning algorithms such as neural networks in the field of dosimetry, especially the analysis of thermoluminescence dosimeter signals, date back several years \citep{Moscovitch1995, Moscovitch1996}. Since then, machine learning methods have not only been used for dose estimation \citep{Moscovitch2006, Theinert.PhD.2018}, but also for the deconvolution of multi-detector signals \citep{Lotfalizadeh2017}, detection of anomalous glow curves \citep{Amit2019, Pathan2020}, and even the generation of artificial glow curves \citep{Kucuk2013, ISIK2020}.
\\\\
In the presented studies, two different types of neural network architectures are used: a fully-connected feed-forward deep neural network (DNN) in \sref{section:Results:DNN} and a convolutional neural network (CNN) in \sref{section:Results:CNN}. Both models are implemented using Keras~\citep{Chollet2015} as the high-level API of Tensorflow 2~\citep{Tensorflow}.
\\
For both architectures, the optimal network and training configurations were determined using a randomized hyperparameter search~\citep{Bergstra2012}.
The performance of the model for each hyperparameter combination is evaluated using a 5-fold cross-validation.
%For this, the networks are trained on 4/5 of the training data and the performance is evaluated on the remaining 20\% of the training data, which is referred to as validation data.
%This process is executed five times until every fold has served as validation set once.
The used performance measure is the mean squared error $\mathrm{MSE}(\Delta t)^2 = \sum_i(t^\mathrm{pred}_\mathrm{pre, i} - t^\mathrm{true}_\mathrm{pre, i})^2$, which is calculated for each validation fold separately and is then averaged over all folds in order to take variations due to the limited training statistics into account.
%A quadratic measure is chosen to penalize outliers.
The resulting network configurations of the best model and preprocessing steps are described in the following.%\todo{ON: Man hätte natürlich schon beschreiben können wie die hyperparameter variiert werden. Muss aber auch nicht.}
%\\
%The uncertainty on the prediction is estimated using the 68\% confidence interval on the prediction error of $\Delta t = t^\mathrm{pred}_\mathrm{pre} - t^\mathrm{true}_\mathrm{pre}$.

\subsubsection{A deep neural network for irradiation date estimation}
The DNN model is trained using the 15 glow curve parameters of n = 548 curves in time space as introduced in \sref{section:Methods:TLDOS} as input for $n = 548$ curves.
All input variables are scaled using a robust standardization, which shifts the median to zero and scales the data such that the interquartile range has a variance of unity. The prediction target $t^\mathrm{true}$ is normalized to a range of [0,1].
The best model of the fully-connected deep neural network has four hidden layers with [64, 128, 38, 16] nodes, which are activated using a tanh function. The output layer with a single node is activated using a linear function. The weights of the connections are adapted by minimizing the mean squared error loss using the Adam optimizer \citep{Kingma.arxiv.2017} with an initial learning rate of $\alpha=0.001$ and a batch size of 32.
The capacity of the DNN is reduced by using a L2-norm regularization \citep{Ng.MLConf.2004} with a strength of $\lambda=0.0005$ to prevent the model from overtraining.

\subsubsection{A convolutional nerual network for irradiation date estimation}
\begin{figure}[h]
	\centering
	\includegraphics[width=0.5\linewidth]{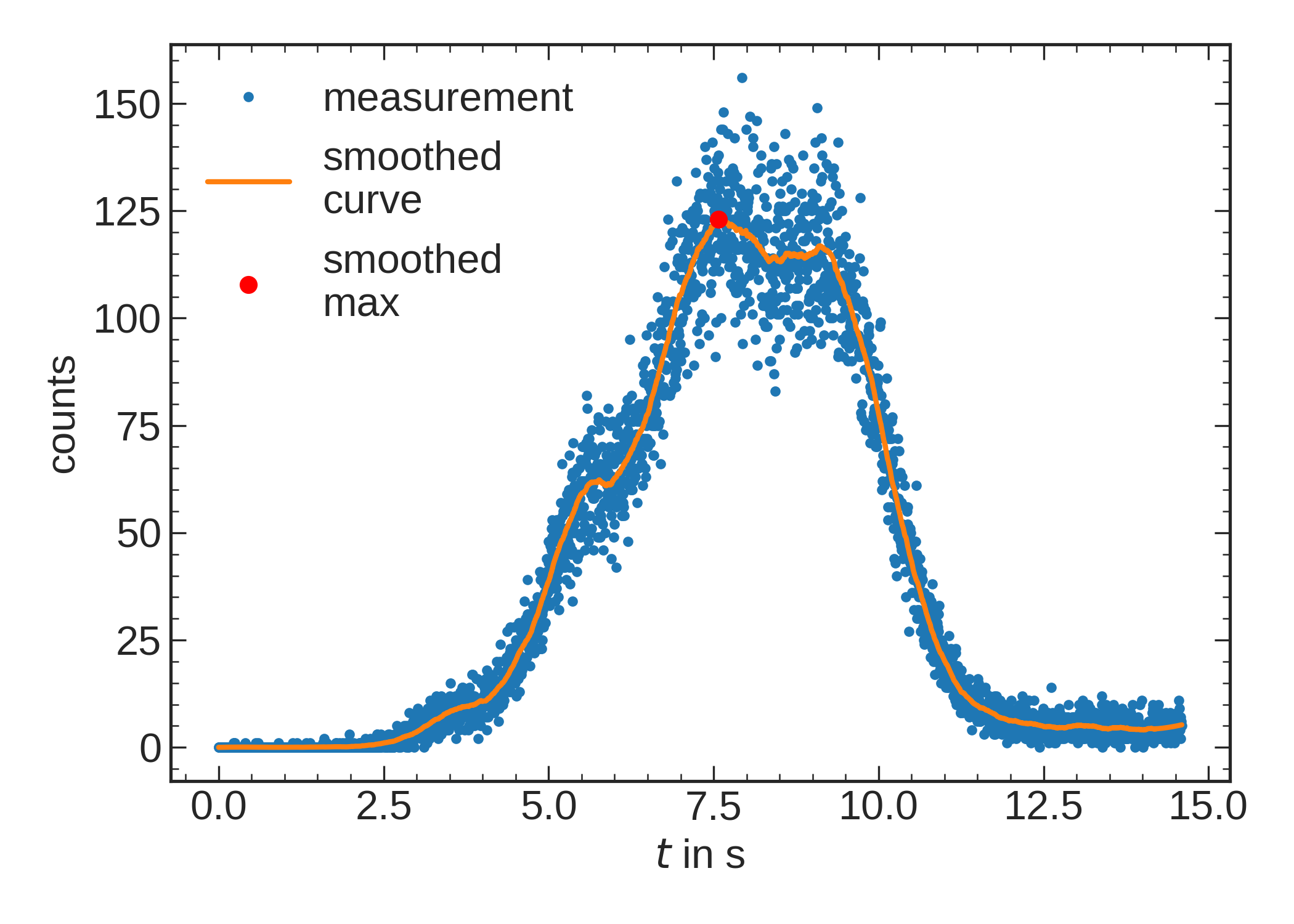}
	\caption{Time resolved photon counts (blue) with a smoothed curve (orange) and its maximum (red), which is used as reference for individual curve normalization.}
	\label{cnn:fig:meas_smoothed_max_normalization}
\end{figure}

\begin{figure}[h]
	\centering
	\includegraphics[width=0.48\linewidth]{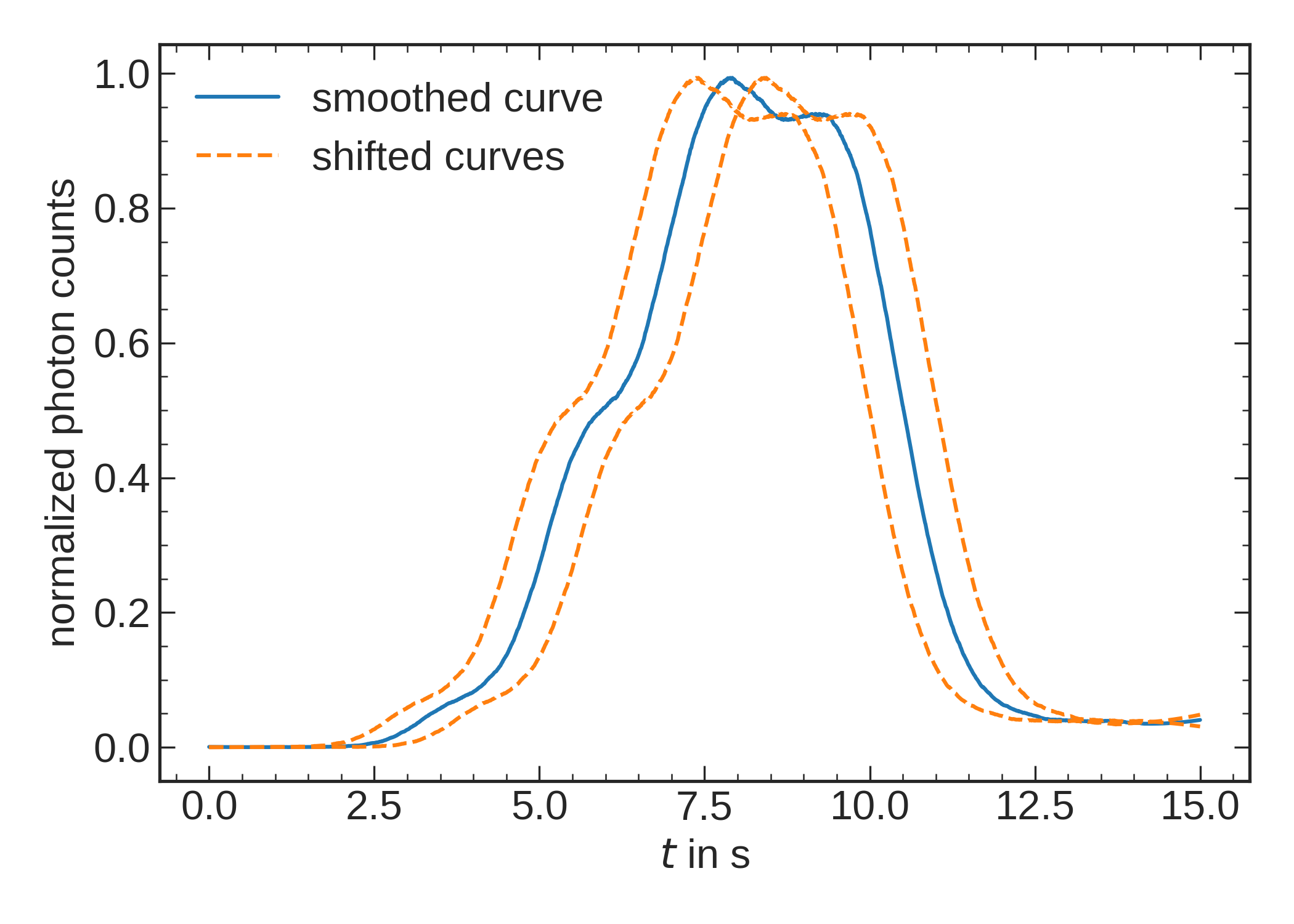}
	\includegraphics[width=0.48\linewidth]{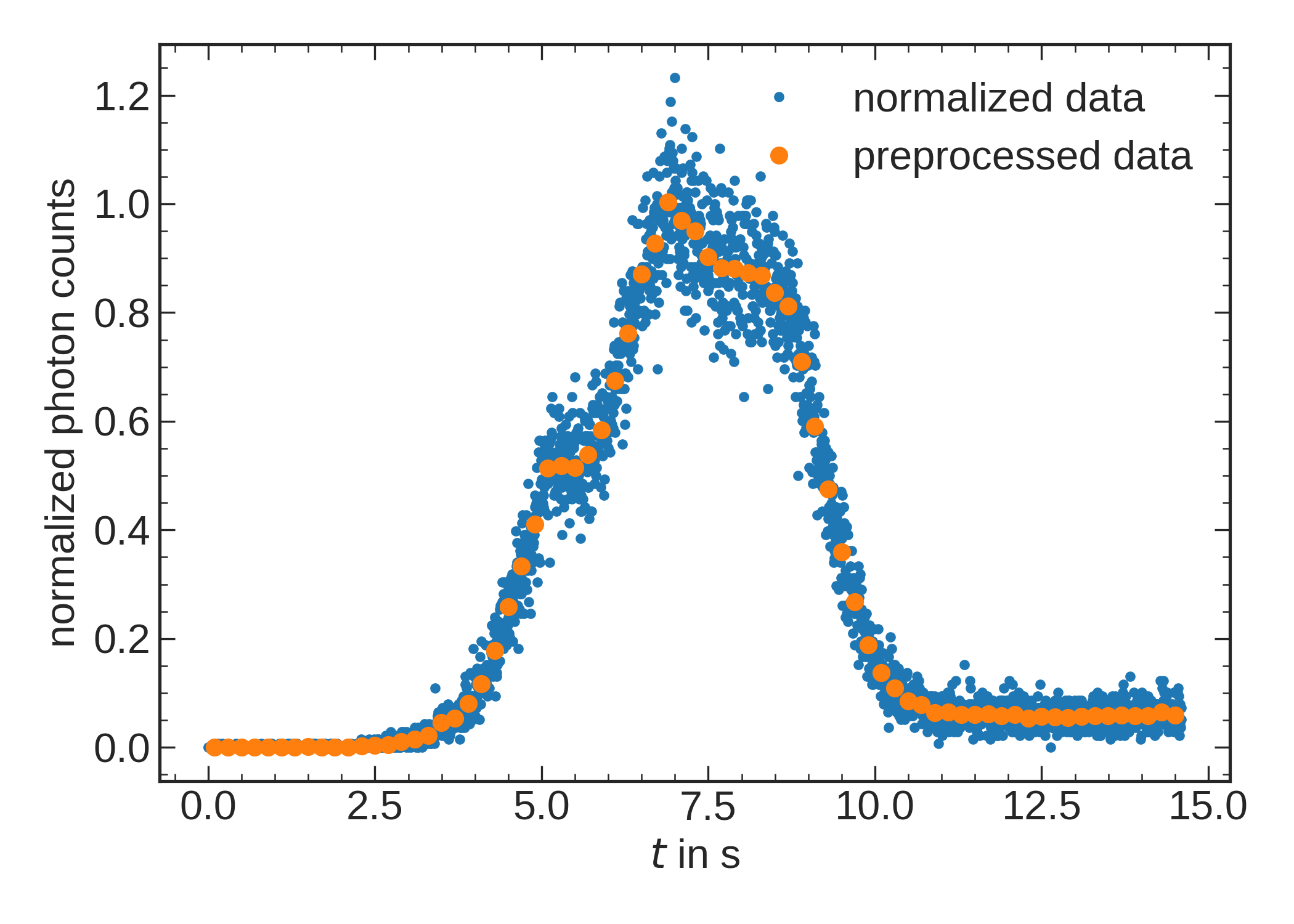}
	\caption{Left: A smoothed glow curve (blue, solid line) and two exemplary shifted and smoothed glow curves (orange, dashed line). Right: Time resolved photon counts (blue) and the output of an average pooling layer with pool size of $k=40$.}
	\label{cnn:fig:shifted_curves}
\end{figure}

\begin{figure}[h]
	\centering
	\includegraphics[width=.8\textwidth]{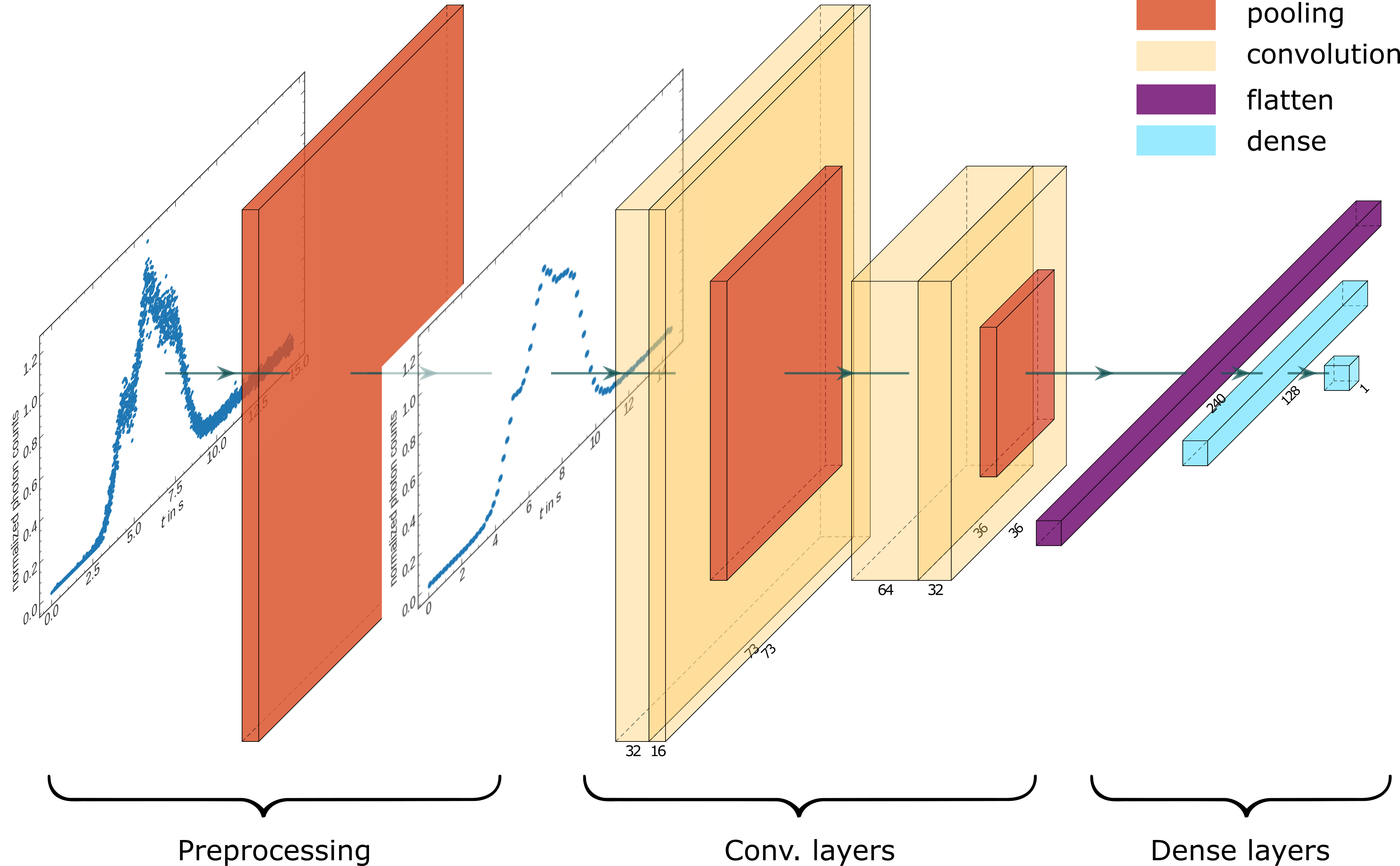}
	\caption{
		Schematic of final network model comprising of pooling layers (red), convolutional layers (yellow), a flatten layer (purple) and two dense layers (blue).
		The input glow curve is shown on the left, the preprocessed glow curve after the first pooling layer.
	}
	\label{cnn:fig:networkarchitecture}
\end{figure}

\begin{figure}[h]
	\centering
	\includegraphics[width=0.48\linewidth]{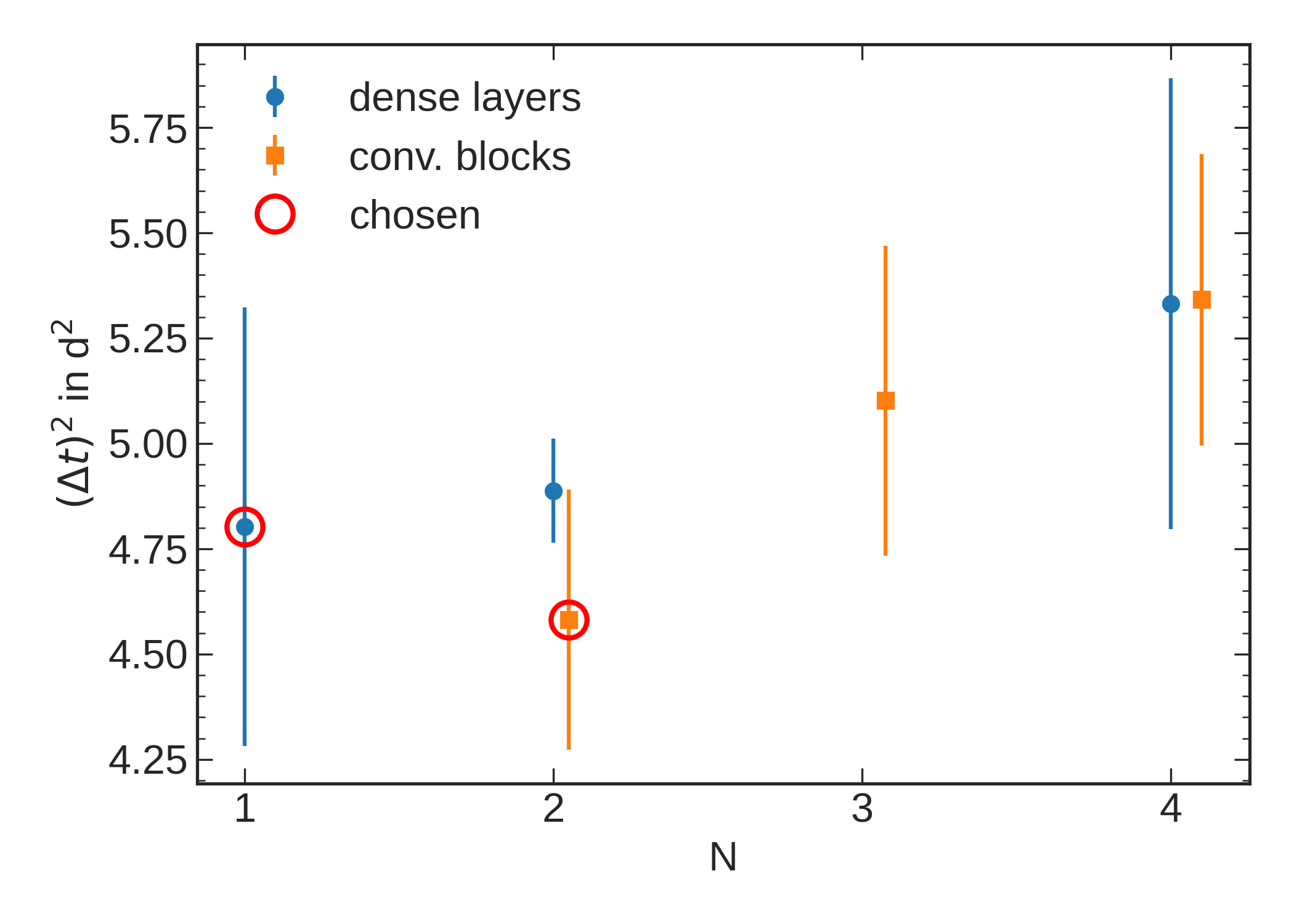}
	\includegraphics[width=0.48\linewidth]{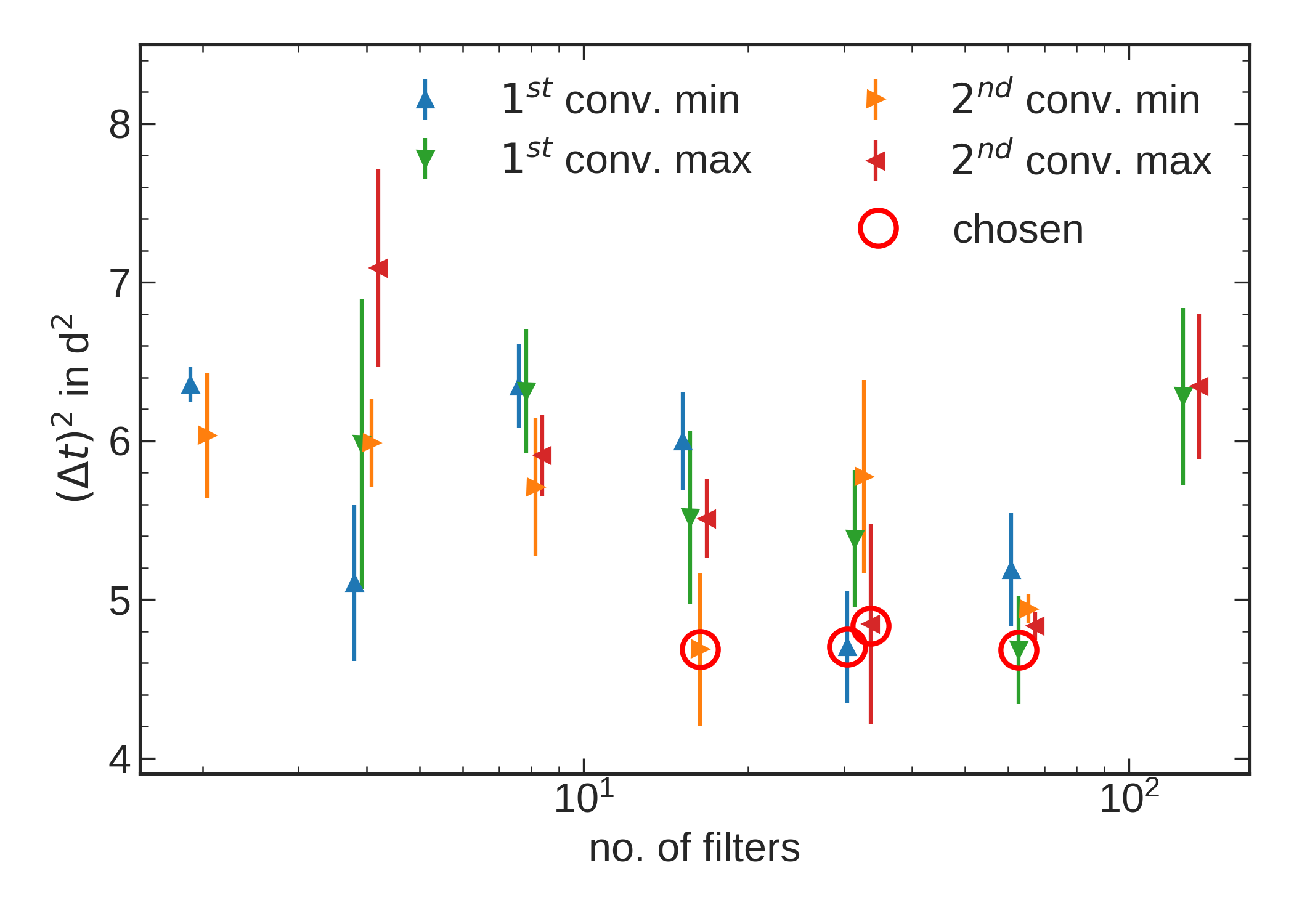}
	\caption{Exemplary results of the grid search.
			The markers denote the average of the five best performing hyperparameter settings including the setting denoted by the tick. The uncertainty shown is the standard deviation of those five results.
			Left: Number of convolutional blocks (orange) and dense layers (blue) against the average validation performance. Right: Minimum and maximum number of filters for the first and second convolution the convolutional blocks.}
	\label{cnn:fig:cnngridsearchdenseconvlayers}
\end{figure}

Convolutional neural networks use trainable filter matrices in order to create features from the data and are known in particular from image processing.
The dimensions of those filters are commonly a lot smaller than the dimension of the original input data and are applied to the data sample by sliding the filter matrix over the data set.
Using one-dimensional convolutions on the glow curves allows the network to automatically learn the optimal filters which create features of the glow curve signal that depend on the irradiation date.
This is expected to yield a better prediction performance because knowledge-based features like those describing a time-resolved glow curve in figure \ref{fig:tl-dos:time-glow-curve} are normally well understandable and interpretable, but not necessarily best suited for the training of a machine learning algorithm for a given task. In addition, the total information content is reduced by using these knowldege-based features and neglecting at the same time the raw time-resolved glow curve data.
\\
For this part of the study, every data sample comprises a pre-irradiation storage time $t_\mathrm{pre}$ (target) and a glow curve (photon count vector) with 2920 entries.
All pre-irradiation storage times $t_\mathrm{pre}$ are normalized to values between 0 and 1.
Several preprocessing steps are applied to the data of the glow curve before the data is given as input to the CNN.
The individual detectors exhibit a varying dose response due to non-constant fabrication conditions resulting in different maxima of the glow curves.
To reduce the impact of the detector responses, each glow curve is normalized individually by dividing them by their respective maximum after application of a smoothing algorithm. The smoothing is performed to prevent measurement noise from affecting the normalization. An exemplary glow curve with the smoothed curve and the normalization point is shown in \fref{cnn:fig:meas_smoothed_max_normalization}.
\\
In addition to the normalization, a data augmentation step is included in the preprocessing to artificially generate a larger data set from the training data samples.
The goal of this is to make the neural network less sensitive to the specific details of individual training data samples. Training on augmented or noisy data instead reduces the probability of the model to just memorize those training samples instead of finding a general model which describes new data equally well.
Only the training data is augmented during the cross-validation while the validation data is kept unchanged in order to obtain a fair performance evaluation. 
The following data augmentation steps are tested as part of a combined hyperparameter optimization as described in \sref{section:Methods:ML}. In the case of the first three, the standard deviation of the Gaussian distributions used for the steps respectively, is varied during the optimization: i) Adding random Gaussian noise to the target $t_\mathrm{pre}$. ii) Adding random Gaussian noise to the individual photon counts $n_{i,j}$. iii) Adding random positional shifts of the glow curve by removing the first or last $k$ entries, sampled as absolute value from a Gaussian distribution, of a curve. In turn, the same number of values is added on the other end of the curve filling up the opposite end containing the average photon counts of the neighbouring glow curve entries. 
A visualization of this process is shown in \fref{cnn:fig:shifted_curves} on the left side. An exemplary glow curve is shown smoothed to allow for better visibility of the effect of shifting the curve left or right.
Finally, the number of generated glow curves using the data augmentation steps on the training data is varied during the hyperparameter search as well.
With regard to the data augmentation, the best performance on the validation data was obtained during the hyperparameter search by creating $n=3000$ augmented curves by performing the curve shifts as described above using random numbers from a Gaussian distribution with mean of 0 and standard deviation of 100. Any other data augmentation step did not lead to better performance on the validation data.	
\\
A final preprocessing step is applied by using an \textit{average pooling layer} to reduce the dimension of the data by calculating the average over a pool size of $k$ photon counts of the glow curve.
The average pooling layer allows for a reduction of the number of photon counts in a single glow curve by averaging over a number of neighbouring count values and makes the model less sensitive to small local variations. As individual photon counts are not necessarily needed for the exposure date estimation, the complexity of the neural network can be reduced and the robustness of the model increased without losing relevant information.
During the hyperparameter search, the best performance on the validation set is achieved by averaging over $k=40$ photon counts confirming this hypothesis.
This reduces the dimension $L$ of the input data from $L = 2920$ to $L = 73$. A resulting curve is exemplarily shown in \fref{cnn:fig:shifted_curves} on the right side.
\\
The final CNN model as found in the hyperparamter search including all layers is shown in \fref{cnn:fig:networkarchitecture}. The average pooling layer of the preprocessing is shown as part of the model.
After the glow curve is passed through the average pooling layer of the final preprocessing step, the model consists out of two convolutional blocks responsible for encoding the glow curve in features which are input to a fully-connected dense layer block for the pre-irradiation stage time $t_\mathrm{pre}$ prediction.
Each of the convolutional blocks consist of two convolutional layers followed by another average pooling layer. The number of convolutional blocks was found to be one of the most sensitive hyperparameter during the search.
\Fref{cnn:fig:cnngridsearchdenseconvlayers} shows two exemplary results of the hyperparameter grid search. The markers denote the average validation performance and their standard deviation of the five best performing hyperparameter settings, where the hyperparameter on the x-axis is kept constant.
Two convolutional blocks are found to be optimal for this model as shown in \fref{cnn:fig:cnngridsearchdenseconvlayers} on the left. The best results were achieved with 32 and 64 convolutional filter in the first block and 16 and 32 convolutional filter in the second block as shown in \fref{cnn:fig:cnngridsearchdenseconvlayers} on the right.
Other hyperparameter of the convolutional layers, like the size of the filter (best=7) and the activation function (best=ReLU) had less impact on the result.
\\
The output of the last convolutional block is flattened into a 1-dimensional array (\fref{cnn:fig:networkarchitecture}, purple) before being passed as input to the dense block of the network. The best results during the hyperparamter search were obtained using only a single dense layer with 256 nodes and ReLU activation function and an output node with linear activation function. A comparison between of the performance with different number of dense layers during the hyperparameter search is shown in \fref{cnn:fig:cnngridsearchdenseconvlayers} on the left side. Other hyperparameter, like the number of nodes in the layer and the activation functions of the nodes had a smaller impact on the result. In order to limit the effective capacity of the model and avoid overtraining, dropout \citep{Srivastava2014} is applied as regularization measure after the flattening and the dense layer. Best results were found to be achieved using a dropout rate of $r=0.4$ for both layers.
\\
The weights of the network are optimized using the Adam optimizer \citep{Kingma.arxiv.2017} with the mean squared error as loss function, an initial learning rate of $\alpha = 1\cdot10^{-4}$ and a batch size of 4, the latter two were varied during the hyperparameter search.

\section{Results\label{section:Results}}
%This section presents the irradiation date estimation using a deep neural network on hand-crafted features derived from the time-resolved glow curve and using a convolutional neural network without manual feature generation.

\subsection{Irradiation date estimation with a deep neural network\label{section:Results:DNN}}

\begin{figure}[h]
	\centering
	\begin{subfigure}[b]{0.48\textwidth}
		\centering
		\includegraphics[width=\textwidth]{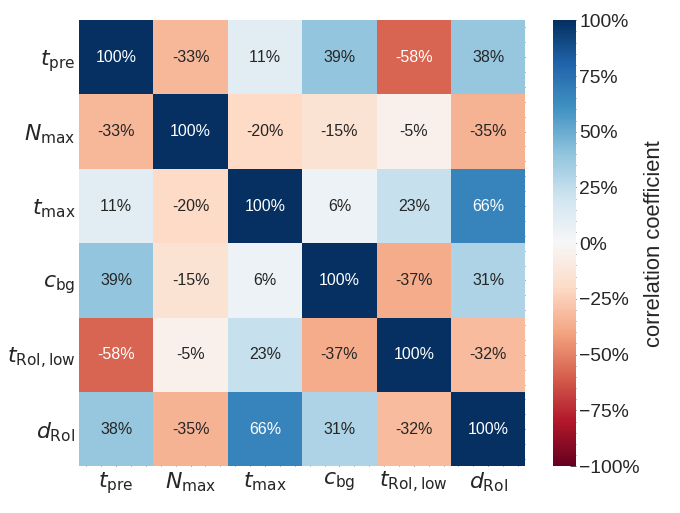}
		\caption{}
		\label{fig:gcana:para_corr}
	\end{subfigure}
	\hfill
	\begin{subfigure}[b]{0.48\textwidth}
		\centering
		\includegraphics[width=\textwidth]{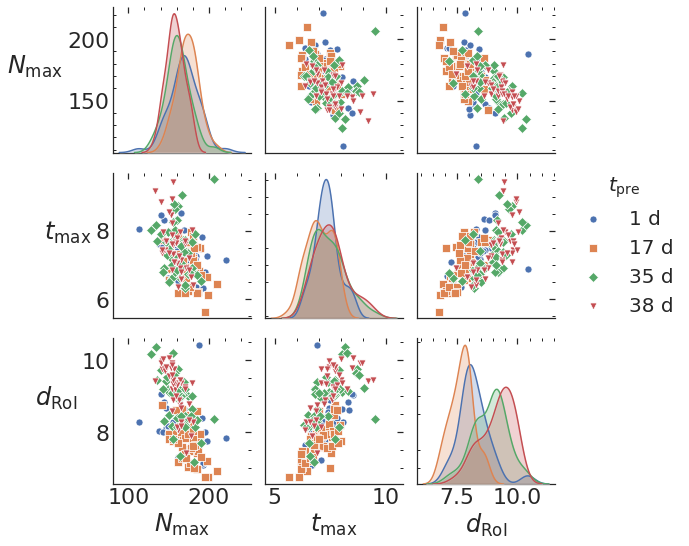}
		\caption{}
		\label{fig:gcana:scatterplot}
	\end{subfigure}
	\caption{(a) Linear correlation coefficients in percentage between different glow curve parameter with the highest correlations towards the fading time. (b) In the non-diagonal the two-dimensional inter-correlations of three selected glow curve parameters are shown in a scatter plot for various pre-irradiation times of $t_\mathrm{pre}= [1, 17, 35, 38]$ days. The diagonal plots show distribution estimations for the individual parameter for the different pre-irradiation times using the kernel density method \citep{Silverman86}, which depict the differences in the distribution of the parameter in dependency of the different pre-irradiation storage times.}
\end{figure}

\begin{figure}[h]
	\centering
	\includegraphics[width=0.6\linewidth]{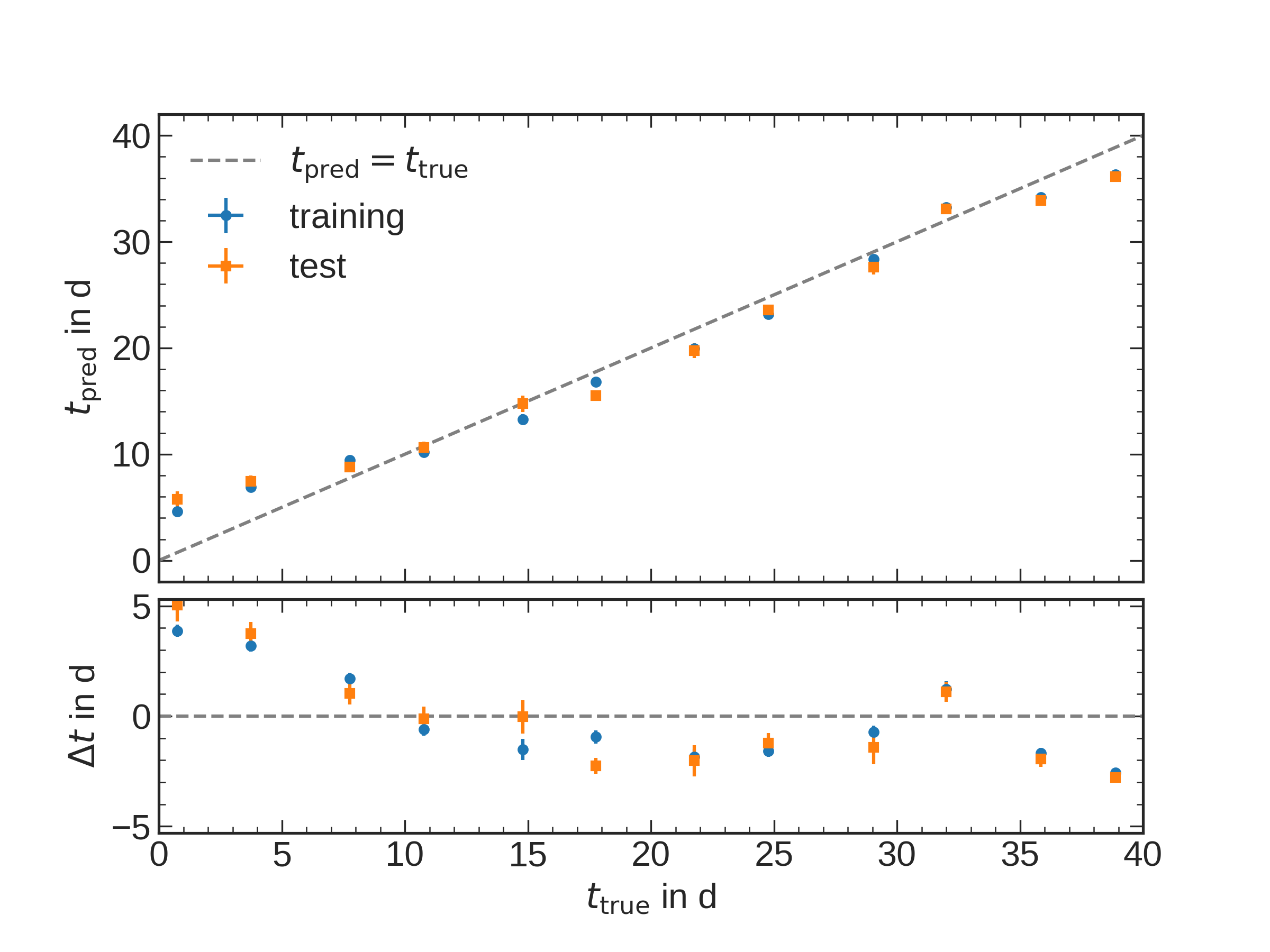}
	\caption{
		The average predicted pre-irradiation storage time per measurement group and its standard error as a function of the true pre-irradiation storage times using the DNN model is shown for training (blue) and test (orange) data.
		The plot below shows the difference $\Delta t = t^\mathrm{pred}_\mathrm{pre} - t^\mathrm{true}_\mathrm{pre}$.
		The uncertainty of the x-axis refers to the averaging of $t^\mathrm{true}_\mathrm{pre}$.
		}
	\label{fig:gcana:NNprediction}
\end{figure}

The five features with the highest linear correlation to the pre-irradiation storage time $t_\mathrm{pre}$ are the maximum number of counts $N_\mathrm{max}$, the time at which the maximum number of counts occurs $t_\mathrm{max}$, the intercept of the background $b_\mathrm{bg}$, the lower limit of the RoI $t_\mathrm{RoI, low}$ and the duration of the RoI $d_\mathrm{RoI}$.
The Pearson correlation coefficients $\rho$ of these parameters to the prediction target $t_\mathrm{pre}$ and between each other are shown in \fref{fig:gcana:para_corr}.
All of these features have a relatively high (anti-)correlation to the predicted value ranging between $(|\rho| = 0.1-0.6)$. Most of them have also relatively high correlations among each other with coefficients varying from larger values of $|\rho| = 0.66$ to smaller values of $|\rho| \approx 0.05$. %, which indicates a decorrelation of the glow curve parameters.
\\
\Fref{fig:gcana:scatterplot} shows in scatter plots in the non-diagonal the inter-correlations of three selected glow curve parameters of various pre-irradiation times of $t_\mathrm{pre}= [1, 17, 35, 38]$ days. The values of one parameter are plotted against the values of another parameter.
The diagonal plots show distribution estimations for the individual parameter for the different pre-irradiation times using the kernel density method \citep{Silverman86}, which depict the differences in the distribution of the parameter in dependency of the different pre-irradiation storage times.
The best discrimination for different pre-irradiation times is obtained with the duration of the RoI, however, there is not a clear linear dependence visible. From the scatter plots it is apparent, that the prediction of the pre-irradiation time is not a simple task and higher-order correlations need to be exploited for this.
\Fref{fig:gcana:NNprediction} shows the mean pre-irradiation storage time and its standard error for each measurement group introduced in \sref{section:Methods:dataset} as a function of the corresponding mean true pre-irradiation storage time together with the relative deviation between the true and the predicted time for both the training and the test data set.
The average predicted pre-irradiation storage times for the different measurement groups do not deviate by more than five days from the average true storage times for neither the training nor the test data. After the first week, which is a realistic time span between the preparation of the detectors and the start of usage by a monitored person, the deviations are even less than three days. The agreement between training and test data performance indicates a good generalization capability of the model meaning it can predict unknown data well and does not memorize training data. 

\subsection{Irradiation date estimation with a convolutional neural networks\label{section:Results:CNN}}

\begin{figure}[h]
	\centering
	\includegraphics[width=0.7\linewidth]{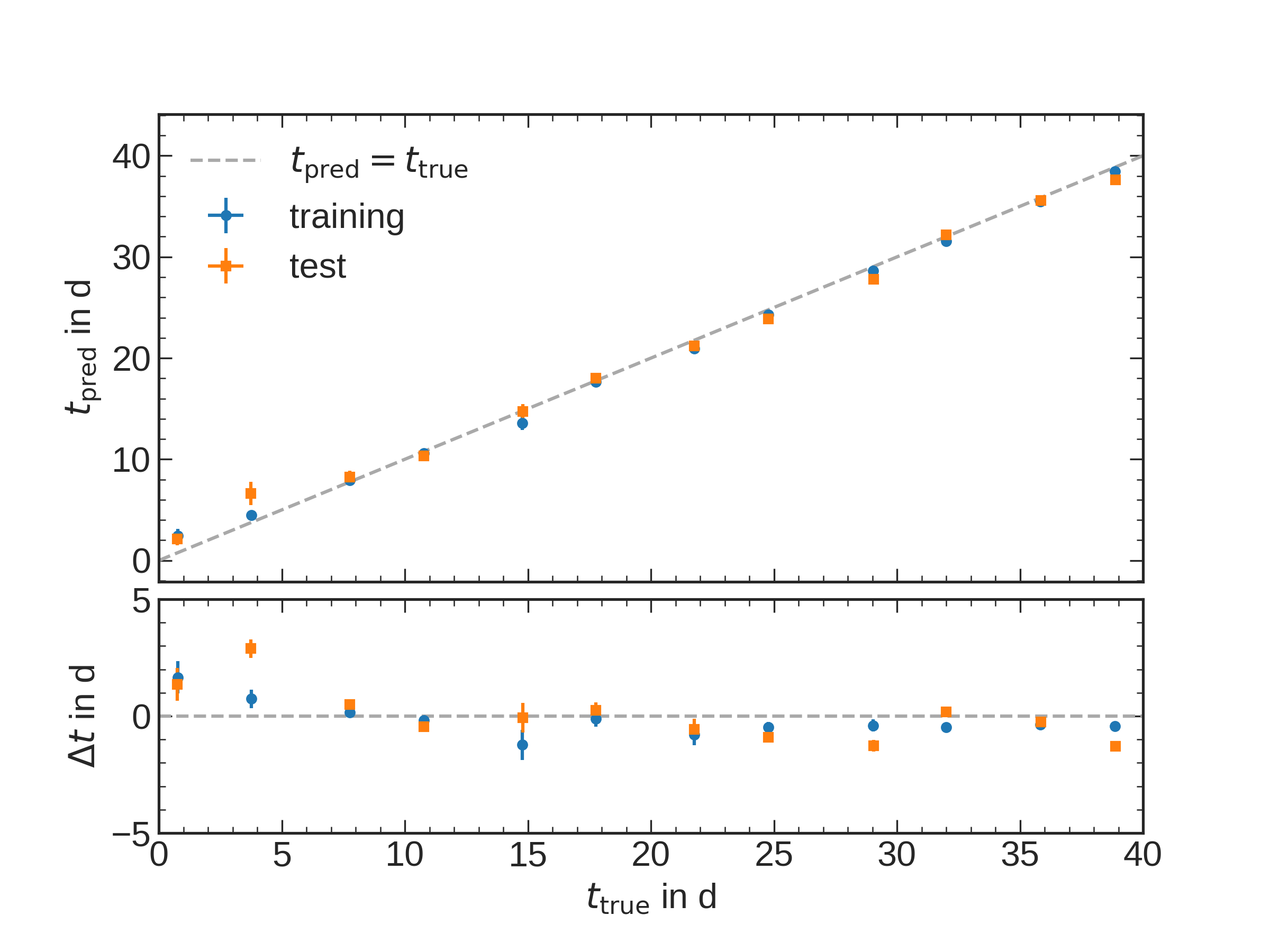}
	\caption{
		The average predicted pre-irradiation storage time per measurement group and its standard error as a function of the true pre-irradiation storage times using the CNN model is shown for training (blue) and test (orange) data.
		The plot below shows the difference $\Delta t = t^\mathrm{pred}_\mathrm{pre} - t^\mathrm{true}_\mathrm{pre}$.
		The uncertainty of the x-axis refers to the averaging of $t^\mathrm{true}_\mathrm{pre}$.
		}
	\label{cnn:fig:ttrue-tpredcomparison}
\end{figure}

The mean predicted pre-irradiation storage time per measurement group of the training and test data samples using the best CNN model is shown in \fref{cnn:fig:ttrue-tpredcomparison} as a function of the true pre-irradiation storage time together with the corresponding standard error.
\\
The mean values per measurement group deviate by less than two days from the true target for almost all pre-irradiation times.
Only the prediction on the test data group with $t_\mathrm{pre} = 4\,$ has a larger deviation from the true value of about three days and deviates also from the prediction of the corresponding training data set. It was found, that this deviation was caused by three outlier predictions of this particular data set.
The standard deviation of the measurement group with $t_\mathrm{pre} = 15\,$d is larger due to a smaller sample size. The predictions on the training data agree in general very well within the uncertainties with the ones on the test data, which indicates that the model is not overtrained.

\section{Discussion\label{section:Discussion}}
\begin{figure}[h]
	\centering
	\includegraphics[width=0.7\linewidth]{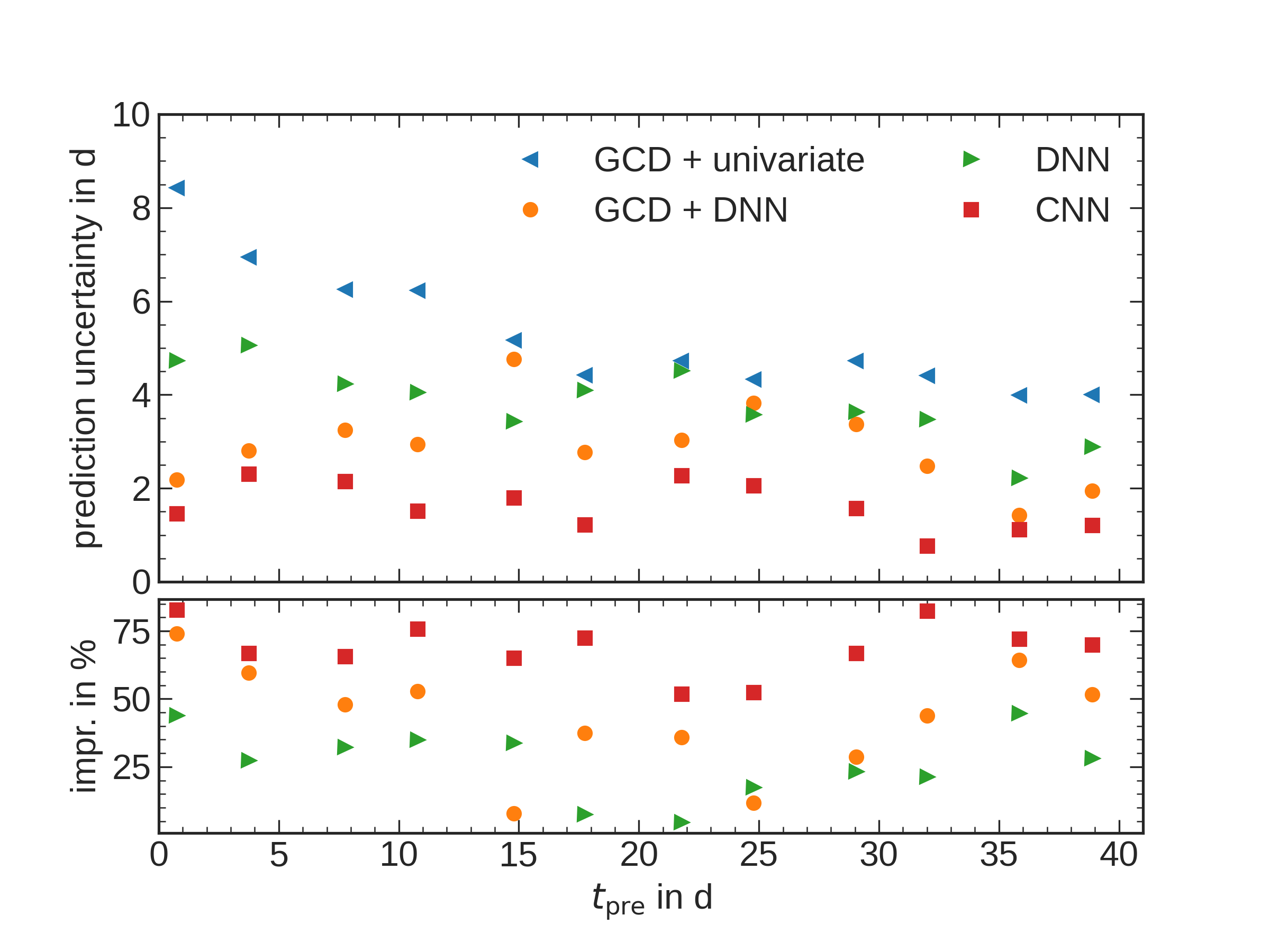}
	\caption{Comparison of prediction uncertainties for $t_\mathrm{pre}$ of the new deep neural network trained with features derived from the time-resolved glow curve (green triangles pointing right) and the convolutional network model red squares) with the previously presented models: univariate analysis of features from glow curve analysis in temperature space (blue triangles pointing left, reproduced from \citep{Mentzel.RadMeas.2020}), deep neural network trained with features from glow curve analysis in temperature space(orange circles, reproduced from \citep{Mentzel.RadMeas.2020}). The plot below shows the respective improvement compared to the univariate baseline model in percent.}
	\label{cnn:fig:convuninncomparison}
\end{figure}

In a prior publication it was demonstrated that using a fully-connected deep neural network (DNN) with multiple glow curve parameters as input obtained from the glow curve deconvolution (GCD) after the temperature reconstruction can significantly improve the estimation precision of the irradiation date compared to a one-dimensional fit approach \citep{Mentzel.RadMeas.2020}. The univariate approach comprised a one-dimensional fit function of a single variable computed from multiple glow curve parameters in dependence of the irradiation date. \Fref{cnn:fig:convuninncomparison} compares the obtained prediction precision of the two earlier published approaches, the univariate using the GCD after the temperature reconstruction (GCD + univariate) and the DNN approach using the GCD after the temperature reconstruction (GCD + DNN) with the two new approaches, the deep neural network without a transformation into the temperature space (DNN) and the convolutional neural network (CNN) approach, which neither uses hand-crafted features nor a GCD.
It shows the 68\% percentile of the absolute deviation $\Delta t = \vert t^\mathrm{pred}_\mathrm{pre} - t^\mathrm{true}_\mathrm{pre} \vert$ as a function of the true pre-irradiation storage time $t_\mathrm{pre}$.
In the subplot below the relative improvement compared to the GCD + univariate approach is shown.
\\
The predictions of the new DNN model yields pre-irradiation dates with uncertainties between 2 and 5 days, which is better than the ones using the GCD + univariate approach but worse than the ones using the GCD + DNN approach. The temperature reconstruction seems to result in glow curves with more physical meaning and hence with features that are more useful to predict the pre-irradiation storage time. In case knowledge-based features are used, the model of the temperature reconstruction and the following deconvolution contains important information, which contributes to the prediction precision and which can not be learned by a DNN from the knowledge-based features in the time space alone.
However, the CNN model is able to predict the pre-irradiation date with higher accuracy compared to all previously presented models over the whole monitoring interval of 42$\,$days.
Most prediction uncertainties are less than two days with an average of $\overline{\Delta t} = (1.6\pm0.1)\,$d over the full monitoring interval. The prediction uncertainty of the test data group with $t_\mathrm{pre} = 4\,$ is not impacted by the three outliers, as discussed earlier, because they are outside of the 68\% percentile.
\\
The CNN model seems to be capable of learning features from the time-resolved glow curves which information is not present in the knowledge-based features of the time space. Furthermore, it seems to make the transformation into temperature space unnecessary, which would allow for much more robust and faster predictions if applied to the dosimetry routine. Being independent of a glow curve model, such as used for the GCD in temperature space, would also allow to apply the approach to a much wider range of irradiation scenarios, e.g. with a combination of different radiative sources or fractionated exposure to irradiation over several days.\\
Having a robust and model-independent estimation of the exposure date in addition to the total received dose has the potential to make passive dosimetry considerably safer.
The improved precision on the irradiation date prediction with an uncertainty of 1-2 days will help the radiation protection officer significantly to identify the event of the irradiation in order to prevent future incidents.
The proposed model serves as a proof-of-concept and is general enough to be applied and extended to other prediction tasks and more realistic scenarios.
\\\\
Currently, the presented CNN model still has several limitations. First of all, the available data statistics is still relatively low for training a deep learning model. Hence data augmentation and strong regularization was necessary to prevent the model from memorizing the data. Similarly, it is not clear at this point if the learned model generalizes well enough to be transferred to a completely independent data set, which is recorded under different conditions. Future studies would strongly benefit from larger data sets.
\\
Furthermore, with the current normalization procedure of the glow curves which aims to be robust against variations in detector sensitivity the model is not capable of predicting the irradiation dose at the same time as any sensitivity to this quantity is effectively removed. This issue however can be solved by using multiple channels in the CNN with different normalization schemes.
\\
Finally, the studies were performed with data recorded at very clean lab conditions, such as a precise irradiation of the detectors with a relatively high dose and the detectors were kept at constant temperature over the whole monitoring interval. More realistic scenarios will result in variations of the data which might impact the result, which could be studied in the future by using data of controlled field studies.

\section{Conclusion\label{section:Conclusion}}

The presented deep convolutional neural network predicts the date of a single irradiation of $D=12\,$mSv with an average uncertainty of approximately 1.6 days within the monitoring interval of
42 days using a TL-DOS thermoluminescence dosimeter.
This surpasses the prediction accuracies of all previously reported algorithms which relied on time-intensive and model-dependent transformations of the glow curve into temperature space with a subsequent deconvolution as well as on manual feature engineering.
The proposed CNN model needs neither of these cumbersome pre-processing steps and is hence not only more accurate but also faster and more robust in the pre-irradiation storage time estimation.
At the same time, the model-independent approach allows for a relatively straight-forward extension to other more complex irradiation scenarios.
This raises the chances of implementing such an analysis step into routine dose monitoring allowing for additional insights into the circumstances of an exposure.
The accurate prediction of important information such as the date or source of exposure can thereby potentially improve the effectiveness of radiation protection measures.
\\

\section{Acknowledgements}
This research was supported by the Deutsche Forschungsgemeinschaft (DFG), project KR 4060/10-1.
%\end{linenumbers}

\newpage
\bibliography{references}

\appendix

\end{document}